\documentclass[11pt,twoside]{article}

\usepackage{asp2006}
\usepackage{epsf}
\usepackage{psfig}
\usepackage{lscape}

\begin{document}
\title{myADS-arXiv - a Tailor-Made, Open Access, Virtual Journal}   %%% Fill in title
\author{E. Henneken, M.J. Kurtz, G. Eichhorn, A. Accomazzi, C.S. Grant, D. Thompson, E. Bohlen, S.S. Murray}   %%% Fill in author names
\affil{Harvard-Smithsonian Center for Astrophysics, 60 Garden Street, Cambridge, MA 02138}    %%% Fill in author affiliations

\begin{abstract} %%% Abstract to run on from here.

The myADS-arXiv service provides the scientific community with a one stop
shop for staying up-to-date with a researcher's field of interest. The
service provides a powerful and unique filter on the enormous amount of
bibliographic information added to the ADS on a daily basis. It also provides a complete view
with the most relevant papers available in the subscriber's field of interest.
With this service, the subscriber will get to know the lastest
developments, popular trends and the most important papers. This makes the
service not only unique from a technical point of view, but also from a
content point of view. On this poster we will argue why myADS-arXiv is a
tailor-made, open access, virtual journal and we will illustrate its unique
character.

The ADS is funded by NASA Grant NNG06GG68G.
\end{abstract}

%%% MAIN BODY OF TEXT GOES HERE. CONSULT "INSTRUCTIONS FOR AUTHORS USING
%%% LATEX2E MARKUP", SECTIONS 2.3-2.6 FOR HELP WITH EQUATIONS, FIGURES,
%%% AND TABLES.

%\section{}   %%% Top level section head (remove "%" symbol)
%\subsection{}   %%% Second level section head (remove "%" symbol)
%\subsubsection{}   %%% Lowest level section head (remove "%" symbol)
%\section*{}    %%% Unnumbered top level section head (remove "%" symbol)
%\subsection*{}   %%% Unnumbered second level section head (remove "%" symbol)

\section{Introduction}

Searching and staying up-to-date with scholarly literature is an essential part of scientific research. With the advent of the World-Wide Web (WWW) and the evolution of electronic publishing, a powerful environment was created to open the vast universe of scientific literature on a world-wide scale. In early 1994 the WWW had become sophisticated enough to allow the search of electronic resources via "web forms". Over the past decade, this environment matured into an unavoidable and indispensable fact of life and tools have emerged in it that have become a crucial ingredient in scientific research. Being able to search vast amounts of data electronically obviously facilitates the review process, but applying advanced technologies such as pattern recognition to the electronic data or by allowing nested searches, one is able to produce results that are unattainable in conventional, non-electronic ways. In that sense one can argue that the available electronic search tools on the WWW even further scientific research.

Using a straight-forward search engine (Google, Yahoo, MSN, AltaVista,...) results in thousands of documents, ranked by some sophisticated algorithm. Even with the advanced versions of these tools, we still find ourselves awash in information. To search the electronic, scholarly literature, scientists need to be able to zoom in on bibliographic data using additional descriptors and search logic. What scholarly tools are available for specialists in astronomy and (astro)physics? The principal bibliographic services are the NASA Astrophysics Data System (NASA ADS), Google Scholar, INSPEC and the Astronomy and Astrophysics Abstracts. Important additional resources are the Science Citation Index (ISI Web of Science), Scopus and ZETOC. Although these tools allow researchers to zoom in on the scholarly literature, they do not offer additional tools to determine the most popular or most cited papers in a given subject. Especially for staying up-to-date, it is essential to be notified of the most popular and most cited papers. Late 2003, the ADS introduced the {\it myADS} service \citep{kurtz03}, a fully customizable newspaper covering (journal) research for astronomy, physics and/or the arXiv e-prints. This service will give the user an overview of the most recent papers by his/her favorite authors, and the most recent, most cited and most popular papers in a particular subject area. Additionally, the user will see an overview of citations to his/her papers. Between a fifth and a quarter of all working astronomers already subscribe to myADS \citep{kurtz05}. {\it myADS-arXiv} is a fully customizable, open access, virtual journal, covering the most important papers of the past week in physics and astronomy. In other words, for the specialist, the myADS-arXiv service provides a one stop shop for staying up-to-date in his/her field of interest.

\section{What is myADS-arXiv?}

myADS-arXiv is based on the existing services of the NASA Astrophysics Data System (see \citet{kurtz00} and Wikipedia) and the arXiv e-print repository (see \citeauthor{ginsparg94} \citeyear{ginsparg94} and \citeyear{ginsparg01}).

The ADS repository is completely synchronized with the arXiv e-prints system. Each night ADS bibliographic records are created for all the e-prints that were newly added. The references are also extracted from the e-prints and matched against existing records in the ADS. Thus we add lists of references to the bibliographic records and we use these references to maintain citation statistics. Both of these elements are used in the myADS-arXiv service. The service also uses readership information from both the ADS and the arXiv to compute the most popular papers in a subject area. By this continuous influx of information, myADS-arXiv provides the subscriber with a service that is as up-to-date as your morning newspaper. The service provides a weekly overview, which offers the unique view on what is happening in a field, and a daily notification. Why is this view unique? It is {\bf Fully Customizable} because you specify the queries that determine the results. myADS-arXiv is an {\bf Open Access} service, because it is {\it totally free} (no subscription costs). Furthermore, it is a {\bf Virtual Journal}, because the overview is a regularly appearing collection of scholarly papers, that are only available in electronic format (the vast majority of e-prints have not yet been published as journal papers). Last but not least, myADS-arXiv covers the {\bf Most Important} papers of the past week. This follows from concordance between e-printed and published papers, and citation statistics: in astronomy and physics, the most important papers are submitted as e-prints first \citep{henneken06}.

The URL for the service is {\bf http://myads.harvard.edu}. From here, you can set up your account and specify the queries for the myADS-arXiv service. A maximum of 2 subject queries and 1 author query can be specified. The results page is like an an automatically generated newsletter. For each of your subject queries, you will get an overview of the "recent", "most popular" and "most cited" papers. The "recent" papers are the newly (i.e. since the previous query) added entries in the e-print database that match your query. The "most popular" papers are found by looking at the also-read statistics for the top 100 of all papers that match your query and the "most cited" papers are obtained from the reference lists of the papers that match your query, published in the previous three months. The link to this newsletter is public, so it can be shared with colleagues. There is also a daily alerting service, showing you the latest e-prints in the categories of your choice, sorted according to a query specified by you (or e-print number if you did not specify one). Entries that match you query are preceded by an asterisk.

\section{Discussion}

Why is myADS-arXiv so unique and powerful? It is because of the combination of the following two factors: the search capabilities of the myADS-arXiv service guarantees the proper selection of bibliographic records for your queries, and the quality of the e-prints guarantees the relevance and importance of the bibliographic records.

The unique machinery that powers the queries consists of {\it reference resolving} (associating reference strings with existing bibliographic records, see \citet{accomazzi99}), {\it bibcode matching} (setting up the e-print/paper concordance), {\it second order operators} (operations on results lists) and {\it also-read statistics} ("people who read paper X, also read paper Y", see \citet{kurtz05}). By associating reference lists in newly added papers with already existing records, we keep citation statistics up-to-date, which allows us to generate lists of "Most Cited Papers". With the reads statistics, we can construct a list of "Most Popular Papers", using the second order operators. The bibcode matching procedure is only relevant in cases where the preprint already appeared in a journal. Titles, abstracts and author lists are indexed, so that they are available for searching.

\begin{figure}[!ht]
  \plottwo{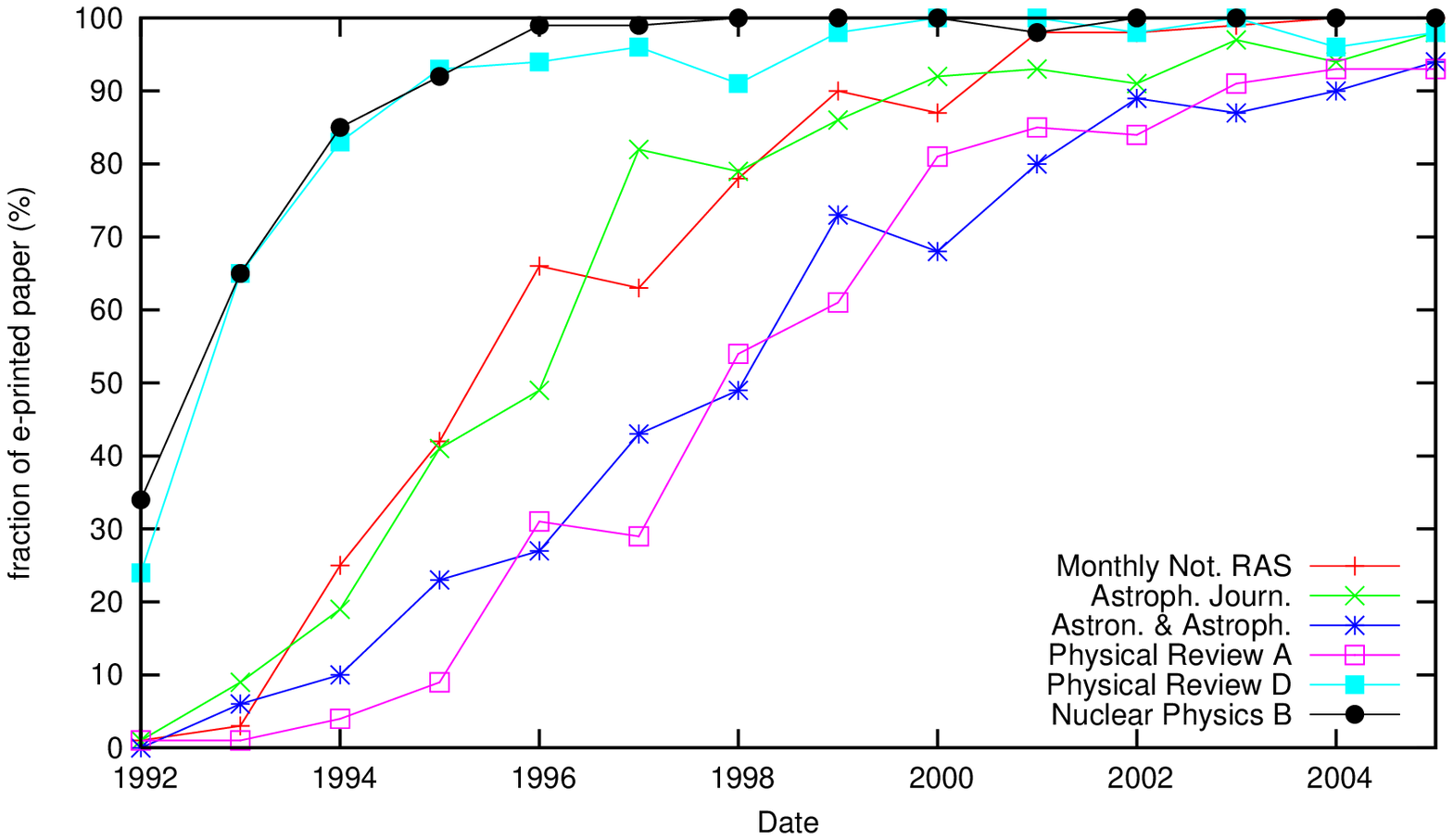}{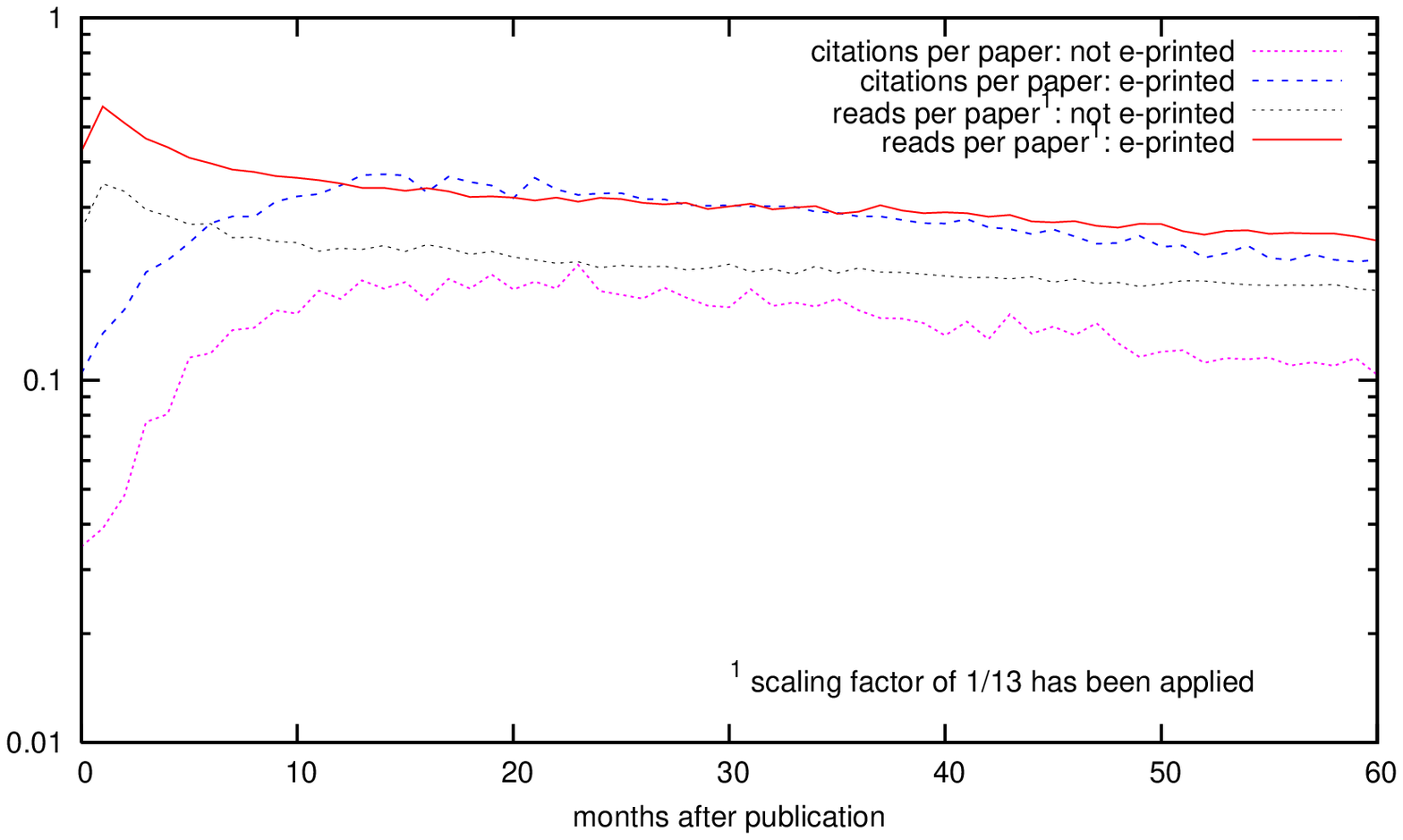}
  \caption{Left: percentage of e-printed papers for the top 100 most cited papers for the journals indicated. Right: citations and reads per paper for ApJ papers (1997-1999)}
\end{figure}

The quality of the e-prints is reflected in the observation: the most important papers in astronomy and physics appear as e-prints on the arXiv. This fact is illustrated by figure 1 (left), which shows for a number of important astronomy and physics journals, the fraction of e-printed papers for the top 100 most cited papers, during the period of 1992 through 2005. Over 90\% appears as e-print first (for Month. Notices RAS and Nucl. Phys. B, it is even 100\%). Just the effect of primacy through early access is not enough to explain the fraction of e-printed papers in the top 100 most cited papers (at a given moment). According to \citet{kurtz05b} there is an effect called "Self-Selection Bias" that results in a further increase of citation rates for e-printed papers. "because papers in the arXiv are not refereed ... this suggests that authors self-censor or self-promote, or that for some reason the most citable authors are also those who first use the new publication venue" \citep{kurtz05b}. The papers are not cited more because they are read more, they are cited more and read more because of their quality. Figure 1 (right) illustrates the fact that e-printed papers are cited and read more than papers that did not appear as e-print.

The rule "better searches give better results" most definitely applies to the myADS-arXiv service. Based on the sophisticated search capabilities of the ADS, myADS-arXiv will provide the most powerful results for those who are able to characterize a field in a couple of keywords and key phrases. Additionally, the use of "simple logic" allows a user to really zoom in on a specific research field.

The myADS-arXiv service is unique in the world of electronic libraries and publishing. No other electronic newsletter or alerting service produces a view on the scholarly physics and astronomy literaturei as comprehensive as this service. From the Harvard-Smithsonian Center for Astrophysics press release in SpaceRef.ca (article number 16658) on the myADS-arXiv service: "It's the best thing since two pieces of sliced bread were assembled to make a sandwich," said Paul Ginsparg, Professor of Physics and Information Science at Cornell University (April 17, 2005).

%\acknowledgements %%% Text of acknowledgements runs on after this command.

%%% THE BIBLIOGRAPHY
%%%
%%% CONSULT SECTION 3 OF "INSTRUCTIONS FOR AUTHORS" FOR HOW TO USE NATBIB.
%%% AUTHORS ARE ENCOURAGED TO USE EITHER THE "THEBIBLIOGRAPY" ENVIRONMENT
%%% BY UNCOMMENTING (DELETING THE "%" SYMBOL) THE COMMANDS BELOW, OR BY
%%% USING THE BIBTEX ENVIRONMENT. TO FIND OUT WHICH IS APPLICABLE TO YOUR
%%% CONTRIBUTION, CONSULT THE VOLUME EDITORS FOR YOUR PROCEEDINGS.
%%%

\end{document}